\documentclass[amsmath,amssymb,10pt]{revtex4}
\pdfoutput=1

\usepackage{multirow}
\usepackage{subfigure}
\usepackage{graphicx}% Include figure files
\usepackage{dcolumn}% Align table columns on decimal point
\usepackage{bm}% bold math

\newcommand{\half}[0]{\frac{1}{2}}

\newcommand{\ld}[0]{\mathcal{L}}

\newcommand{\qsubrm}[2]{{#1}_{\textrm{#2}}}

\newcommand{\rbm}[1]{{\bf{#1}}}

\newcommand{\ci}[0]{{i}}

\newcommand{\sech}[0]{\textrm{ sech}}
\newcommand{\fref}[1]{{Fig. \ref{#1}}}

\let\oldsqrt\sqrt
\def\sqrt{\mathpalette\DHLhksqrt}
\def\DHLhksqrt#1#2{%
\setbox0=\hbox{$#1\oldsqrt{#2\,}$}\dimen0=\ht0
\advance\dimen0-0.2\ht0
\setbox2=\hbox{\vrule height\ht0 depth -\dimen0}%
{\box0\lower0.4pt\box2}}

\begin{document}

\title{Formation and evolution of kinky vortons}
\author{Richard A. Battye}
\email{richard.battye@manchester.ac.uk}
\affiliation{Jodrell Bank Centre for Astrophysics, School of Physics and Astronomy, The University of Manchester, Manchester M13 9PL, U.K}
\author{Jonathan A. Pearson}
\email{jp@jb.man.ac.uk}
\affiliation{Jodrell Bank Centre for Astrophysics, School of Physics and Astronomy, The University of Manchester, Manchester M13 9PL, U.K}
\author{Paul M. Sutcliffe}
\email{p.m.sutcliffe@durham.ac.uk}
\affiliation{Department of Mathematical Sciences, Durham University, Durham DH1 3LE, U.K}
\author{Simon Pike}
\affiliation{Jodrell Bank Centre for Astrophysics, School of Physics and Astronomy, The University of Manchester, Manchester M13 9PL, U.K}
\date{\today}
\begin{abstract}
We present field theory simulations of a model with $\mathbb{Z}_2\times U(1)$ symmetry in (2+1)-dimensions. This model has two discrete vacua, allowing for domain walls, and also a conserved Noether charge. For initial conditions in which the field is placed randomly in one of the vacua and given a homogeneous background charge, we find that the number of walls does not scale in the standard way. We argue that the Noether charge and current become localized on the walls, forming kinky vortons, (2+1)-dimensional analogues of cosmic vortons. These loops of wall can be long lived, or even stable, depending on their precise characteristics. We suggest that our simulations illustrate a possible mechanism for dynamical frustration of domain wall networks and that cosmic vortons will form naturally in $U(1)\times U(1)$ models.
\end{abstract}

 \maketitle

\section{Introduction}
Domain walls are field theoretic solutions which interpolate between discrete vacua. They can be of interest in the context of particle physics and cosmology where the field is relativistic and fundamental~\cite{VS}, or in condensed matter where the field is non-relativistic and represents an order parameter of the system~\cite{Thouless}. Once formed at a phase transition, the number of walls is expected to scale $\propto t^{-1}$ in relativistic systems~\cite{Hindmarsh1, Avelino1, Avelino2,Avelino3, battye_moss_o3} which will be the focus of the present paper; essentially a network will lose energy as quickly as possible within the bounds imposed by causality and the causal horizon is the only length-scale in the problem. Similar standard results exist for non-relativistic systems~\cite{Bray}. 

Cosmic vortons are loops of superconducting string~\cite{witten} which are stabilized by the existence of charge and current which form a condensate coupled to the vortex field~\cite{DavisShellard}. Recent numerical work has investigated the stability of such configurations in (3+1)-dimensions~\cite{LempShell1, LempShell2,BattyeSut1,RaduVolk}, but due to the large variety of length-scales in the problem, this has only allowed quantitative validation of the standard picture for their formation and evolution in a small range of parameter space. 

Kinky vortons are (2+1)-dimensional analogues of cosmic vortons which replace the superconducting vortex with a superconducting domain wall~\cite{BattyeSut2} (or kink). In this lower dimensional case extensive numerical investigations have been possible. In addition due to the fact that exact analytic solutions are possible for the superconducting condensate, independent of the charge and current,  this has allowed analytic investigation of their stability properties in the thin ring limit~\cite{BattyeSut3} developed by Carter and collaborators~\cite{Carter1, Carter2, Carter3, Carter4, Carter5, Carter6, Carter7}. This work appears to suggest that the basic tenets of cosmic vorton evolution, that they might eventually come to dominate the energy density of the Universe, are correct (see, for example, refs.~\cite{BrandenburgerCarterDavisTrodden, MartinsShellard3, MartinsShellard4}).

In this paper we will investigate the evolution of networks of these superconducting domain walls formed from random initial conditions coupled to a homogeneous background charge. In particular we will investigate the scaling properties of the number of walls as a function of time. We find that the effects of the conserved charge and current, which localize on the walls, appear to slow down the natural propensity of kinky-loops to collapse under their own tension, violating the standard law of scaling. Some of the loops appear to have similar properties to stable kinky vortons suggesting that such objects can form rather naturally and, although we cannot confirm their absolute stability due to our limited dynamical range, they are, at the very least, long-lived.

The scaling properties of domain wall networks have another interesting cosmological application. It has been suggested that they may form a frustrated lattice which could be a candidate for a dark energy component with $w=P/\rho=-2/3$~\cite{BucherSpergel, BattyeBucherSpergel}. However, all simulations of domain wall networks, even those with junctions (for example refs.~\cite{Avelino1, Avelino2,Avelino3, battye_moss_o3}), appear to be compatible with the standard scaling law - although they all have relatively poor dynamical range compared to cosmological scales. In a separate paper \cite{BattyePearson} we will investigate how similar models to those considered here, which also allow for junctions, behave in an analogous way, but with some extra complications.

\section{Model and Numerical Implementation}

The (2+1)-dimensional model we will investigate is described by the Lagrangian density
\begin{eqnarray}
\ld = \partial^{\mu}\phi\partial_{\mu}\phi +  \partial^{\mu}\sigma\partial_{\mu}\bar{\sigma} - \frac{\lambda_{\phi}}{4}\left( \phi^2 - \eta_{\phi}^2 \right)^2-\frac{\lambda_{\sigma}}{4}\left( | \sigma |^2 - \eta_{\sigma}^2 \right)^2 - \beta\phi^2|\sigma|^2,
\end{eqnarray}
which has a global $\mathbb{Z}_2\times U(1)$ symmetry. Throughout the subsequent discussion we fix the model parameters $\lambda_{\phi} = \lambda_{\sigma} = 2, \beta=1$; see ref. ~\cite{BattyeSut2} for a discussion of parameter values in this theory, and the motivation for this particular choice. The real scalar field $\phi(t,\rbm{x})$ is  the broken domain wall-forming field and the complex scalar field $\sigma(t,\rbm{x})$ will act as the superconducting  condensate.
 
The equations of motion admit a $y$-directed domain wall solution~\cite{BattyeSut2}
\begin{subequations}
\label{eq:Sec1:solns}
\begin{eqnarray}
\label{eq:sec1:soln-phi}
\phi &=& \eta_{\phi}\tanh\left(m_{\phi}x\sqrt{1-\alpha - \hat{\chi}} \right),\\
\label{eq:sec1:soln-sigma}
\sigma &=& |\sigma|e^{\ci(ky+\omega t)},\\
|\sigma| &=& \sqrt{2}\eta_{\phi}\sqrt{\alpha - \half + \hat{\chi}}\sech\left(m_{\phi}x\sqrt{1-\alpha - \hat{\chi}} \right),
\end{eqnarray}
\end{subequations}
where we have defined $\alpha \equiv \left({\eta_{\sigma}}/{\eta_{\phi}}\right)^2, \hat{\chi} \equiv{\chi}/{m_{\phi}^2}, \chi \equiv \omega^2-k^2$,
\begin{eqnarray*}
m_{\phi}^2 = \frac{\lambda_{\phi}}{2}\eta_{\phi}^2,\quad m_{\sigma}^2 = \frac{\lambda_{\sigma}}{2}\eta_{\sigma}^2+\chi.
\end{eqnarray*}
Existence of a solution requires that $1/2 - \alpha < \hat{\chi} < 1-\alpha$ and this solution explicitly exhibits many of the expected properties of superconducting solitons such as current quenching.

One can use this solution to construct ring solutions if the radius, $R$, is sufficiently large that corrections due to the curvature are negligible. Such kinky vorton solutions can be specified by the winding number $N$ with $k = N/R$, and the conserved charge of the condensate field is given by
\begin{eqnarray}
Q = \int d^2x\,\qsubrm{\rho}{Q} = \frac{1}{2\ci}\int{d}^2x\left( \dot{\sigma}\bar{\sigma} - \dot{\bar{\sigma}}\sigma\right).
\end{eqnarray}
The current is given by
\begin{eqnarray}
\rbm{J} = \frac{1}{2i}\left( \bar{\sigma}\nabla\sigma - \sigma\nabla\bar{\sigma}\right).
\end{eqnarray}
The stable radius $R_{\star}$ can be computed for sufficiently large values of $N$ and $Q$~\cite{BattyeSut2}.

We evolve the equations of motion in (2+1)-dimensions by discretizing on a regular square grid of $P^2$ points and grid-spacing $\Delta x$. The time evolution is performed using a second-order leapfrog algorithm, with time-step size $\Delta t$ and  spatial derivatives are discretized to fourth order. Typically we will use $P=4096$ and $\Delta x = 0.5, \Delta t = 0.1$, although some simulations have used $P=1024$ in order to reduce computational requirements. All simulations presented here use $\eta_{\phi} = 1, \eta_{\sigma} = \sqrt{3}/2$ (such that $\alpha = 3/4$), but similar results are found for all values of $\alpha$ compatible with the existence of a kinky vorton solution. We use periodic boundary conditions, which sets a light-crossing time: the simulation time taken for an emitted signal to interact with itself after passing through the periodic boundaries, given by $\tau = \half P\Delta x$. Our results on the scaling of the number of walls are only valid upto this light-crossing time; however, structures that persist after this time are still of interest, particularly when they appear to be stable.

Initial conditions are chosen in order to represent a phase transition in the $\phi$-field and a condensate with a uniform charge density in the $\sigma$-field. We note that it is not our intention to model a realistic phase transition by which kinky vortons might be formed, rather to do something which is random and relatively painless from a practical point of view. At each grid-point the $\phi$-field is placed randomly into one of the two points on the vacuum manifold, with $\dot{\phi}=0$. The $\sigma$-field is set to $\sigma = A e^{\ci \omega t}$ where $A$ and $\omega$ are real constants for which the initial charge density is $\qsubrm{\rho}{Q}(0) = A^2\omega$. All simulations presented here  have $\omega = 1.0$, and $A$ is varied to change the initial charge density; we have checked that similar results are found if we fix $A$ and vary $\omega$. Although these initial conditions are random, they will contain unphysical energy gradients between adjacent grid-points. In order to ameliorate this problem we introduce a damping term which only acts on the $\phi$-field for the first 200 time-steps in our simulation. This results in a reduced energy and aids condensation into smooth domains, without reducing the total charge which remains approximately constant.

In order to study the scaling properties we need a method which can be used to estimate the number of walls in the simulation from the fields. We do this by first deciding which of the two minima $\phi$ is nearest at each grid point. If either of the adjacent grid points (only above and to the right to avoid duplication) return different minima then the wall count is incremented.

\section{Results}
 \begin {figure}[!ht]
      \begin{center}
{\includegraphics[scale=0.9]{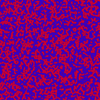}\quad\includegraphics[scale=0.9]{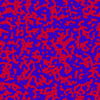}\quad\includegraphics[scale=0.9]{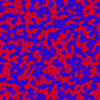}\quad\includegraphics[scale=0.9]{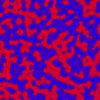}\quad\includegraphics[scale=0.9]{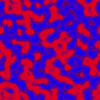}}\\\vspace{0.5cm}

{\includegraphics[scale=0.9]{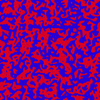}\quad\includegraphics[scale=0.9]{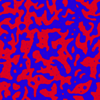}\quad\includegraphics[scale=0.9]{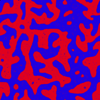}\quad\includegraphics[scale=0.9]{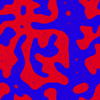}\quad\includegraphics[scale=0.9]{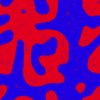}}\\\vspace{0.5cm}

{\includegraphics[scale=0.9]{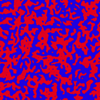}\quad\includegraphics[scale=0.9]{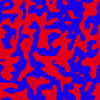}\quad\includegraphics[scale=0.9]{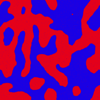}\quad\includegraphics[scale=0.9]{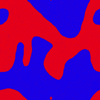}\quad\includegraphics[scale=0.9]{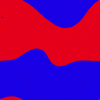}}\\\vspace{0.5cm}

{\includegraphics[scale=0.9]{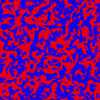}\quad\includegraphics[scale=0.9]{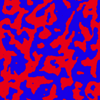}\quad\includegraphics[scale=0.9]{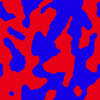}\quad\includegraphics[scale=0.9]{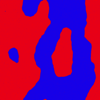}\quad\includegraphics[scale=0.9]{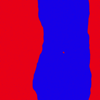}}
      \end{center}
\caption{Evolution of the $\phi$-field for $\qsubrm{\rho}{Q}(0)=0.25, 0.09, 0.01, 0$ (from top to bottom). The colours correspond to each of the two points on the vacuum manifold (red/light: $\phi = +1$; blue/dark: $\phi = -1$).  The images are at $t = 80, 160, 320, 640, 1280$ (left to right) and we have used $P=4096$. There is a clearly visible difference in the long-term behaviour of the low charge cases $\qsubrm{\rho}{Q}(0)=0$ and 0.01, and the higher charge cases $\qsubrm{\rho}{Q}(0)=0.09$ and 0.25.}
\label{fig:fieldevol}
\end {figure}       

\begin {figure}[!ht]
      \begin{center}
	\input{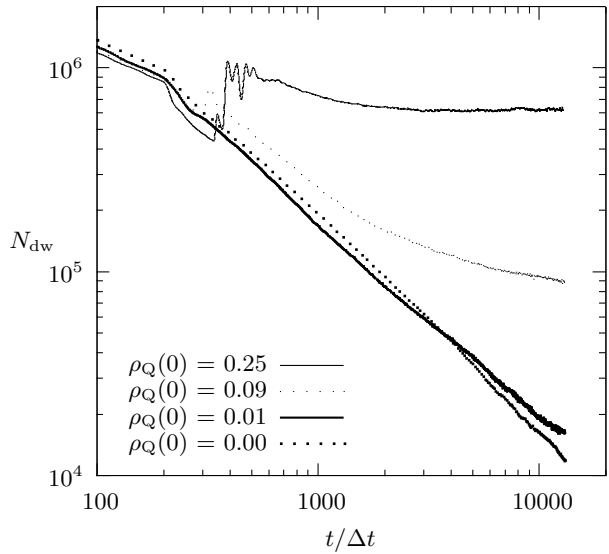}
      \end{center}
\caption{Evolution of the number of domain walls for various $\qsubrm{\rho}{Q}(0)$ with $P=4096$. For $\qsubrm{\rho}{Q}(0)=0$ and 0.01 we see that the number of walls appears to scale $\propto t^{-1}$, whereas the evolution is modified for large values of $\qsubrm{\rho}{Q}(0)$.}\label{fig:4096-nwalls-varyalpha}
\end {figure}  
 
\begin {figure}[]
      \begin{center}
{\includegraphics[scale=0.9]{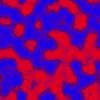}\quad\includegraphics[scale=0.9]{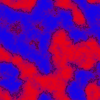}\quad\includegraphics[scale=0.9]{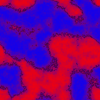}\quad\includegraphics[scale=0.9]{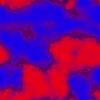}}\\\vspace{0.5cm}
{\includegraphics[scale=0.9]{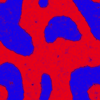}\quad\includegraphics[scale=0.9]{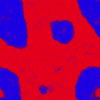}\quad\includegraphics[scale=0.9]{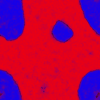}\quad\includegraphics[scale=0.9]{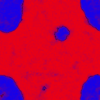}}
      \end{center}
\caption{Evolution of the $\phi$-field for $\qsubrm{\rho}{Q}(0)=0.25$ (top), $0.09$ (bottom), at $t = \tau, 2\tau, 3\tau, 4\tau$ (from left to right). Due to dynamical range constraints we have used $P=1024$ for these simulations. We see that there is very little evolution beyond the light crossing time, suggesting that the structures have frozen in.}
\label{fig:freeze}
\end {figure} 

\begin {figure}[]
      \begin{center}
\subfigure[$\,\qsubrm{\rho}{PE}$]{\includegraphics[scale=0.7]{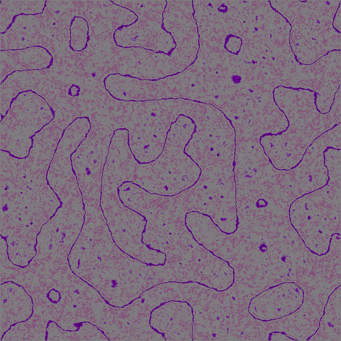}}\quad
\subfigure[$\,\Re(\sigma)$]{\includegraphics[scale=0.7]{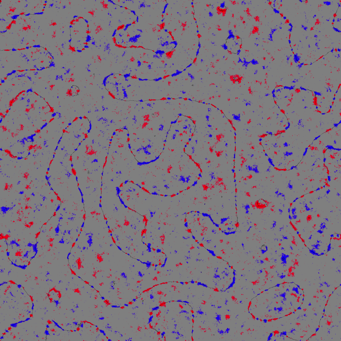}}\\\vspace{0.3cm}
\subfigure[$\,\qsubrm{\rho}{Q}$]{\includegraphics[scale=0.7]{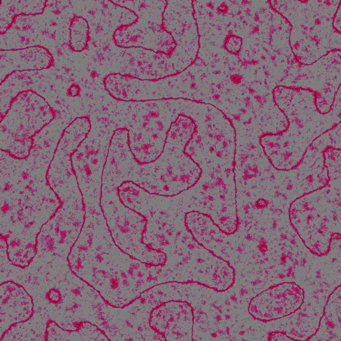}}\quad
\subfigure[$\,|\rbm{J}|^2$]{\includegraphics[scale=0.7]{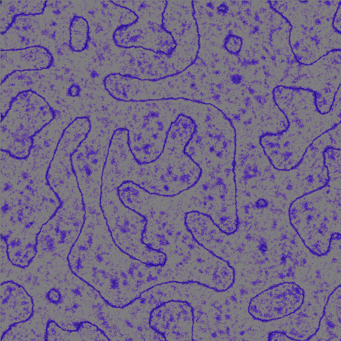}}
      \end{center}
\caption{Results of the simulation with $\qsubrm{\rho}{Q}(0) = 0.09$ at $t = 640$ : (a) potential energy density, $\rho_{\rm PE}$ using colouring routine where purple is the maximum value, grey is where less than 10\% of its maximal value; (b) $\Re(\sigma)$ where  red/light and blue/dark denotes $\Re(\sigma)$ positive/negative, and grey where $\Re(\sigma)<40\%$ maximum value; (c) charge density, $\rho_{\rm Q}$,  where red denotes positive, grey denotes $\qsubrm{\rho}{Q} < 10\%$ maximal value; (d) the modulus of the current $|\rbm{J}|^2$ with blue denoting $|\rbm{J}|^2$, grey denotes $|\rbm{J}|^2<2\%$ maximal value. All images are coloured on a gradient until the specified threshold. Counting loops reveals one large loop and a number of other smaller loops none of which wrap around the toroidal geometry. It is clear that $\rho_{\rm PE}$, $\qsubrm{\rho}{Q}$ and $|{\bf J}|^2$ are localized on the domain walls, and that $\Re{(\sigma)}$ oscillates between positive and negative values as one traverses a loop of wall.}
\label{fig:quant}
\end {figure}  

\begin{figure}[]
  \centering
  \begin{minipage}[c]{0.25\textwidth}
    \centering
\begin{tabular}{||c|c|cc||}
\hline
&&\multicolumn{2}{c||}{\textbf{Time} }\\
\hline\hline
Loop &&640 & 1280 \\
\hline\hline
\multirow{3}{*}{\textbf{1}} 
&$N$ & 7 & 3 \\
&$Q$&2300&1700 \\
& $R$& 77 & 40 \\
&$R_{\star}$& 51 & 29 \\
\hline
\multirow{3}{*}{\textbf{2}} 
&$N$ & 35 & 27 \\
&$Q$&14000&12000 \\
& $R$& 280 & 260 \\
&$R_{\star}$& 282 & 229 \\
\hline
\multirow{3}{*}{\textbf{3}} 
&$N$ & 11 & 4 \\
&$Q$&3000&3400 \\
& $R$ & 130 & 100\\
&$R_{\star}$& 74 & 47 \\
\hline
\hline
\end{tabular}
  \end{minipage}
  \begin{minipage}[c]{0.73\textwidth}
\subfigure[ $\,t=640$]{\includegraphics[scale=0.5]{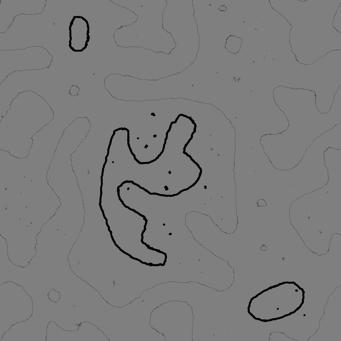}}\quad
\subfigure[ $\,t=1280$]{\includegraphics[scale=0.5]{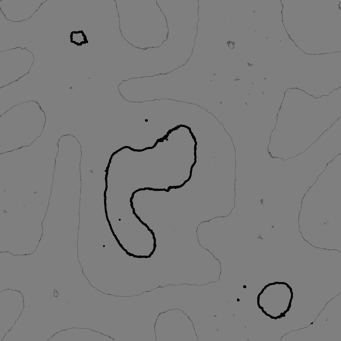}}
  \end{minipage}
\caption{Properties of three of the loops in the $\qsubrm{\rho}{Q}(0) = 0.09$ simulation. $N$ is found by counting the number of red-blue pairs along a loop; $Q$ is computed by integrating the charge density over a region 10 grid-squares from the loop; and $R$ is computed from the area of the loop. $R_{\star}$ is the radius of a stable kinky vorton with the given $N$ and $Q$. The highlighted loops are those for which the table gives the properties. The loops are labelled from top left to bottom right (1-3). The fraction of the charge which resides on all loops to the total charge in the box (i.e. not just those highlighted) is $7\%$ and $6\%$ for the two time-steps shown.}
\label{fig:kinky}
\end{figure}

We have peformed simulations with $P=4096$ upto $t=\tau$ for four different values of $A$ which correspond to initial values of $\qsubrm{\rho}{Q}(0)=0.25$, 0.09, 0.01 and 0. The evolution of the fields are presented in Fig.~\ref{fig:fieldevol} at times which are multiples of 2 from $t= 80$ upto 1280 (which is just beyond $t=\tau$). If the network is scaling, then one would expect the size of the domains to increase by a factor of 2 from one plot to the next. The first thing to note is that the zero and low charge case ($\qsubrm{\rho}{Q}(0)=0.01$) look very similar, whereas the higher charge cases ($\qsubrm{\rho}{Q}(0)=0.09$ and $\qsubrm{\rho}{Q}(0)=0.25$) are qualitatively different, with much smaller domains particularly in the case of $\qsubrm{\rho}{Q}(0)=0.25$. The number of walls as a function of time are presented in \fref{fig:4096-nwalls-varyalpha}. This confirms our observations about \fref{fig:fieldevol}: for $\qsubrm{\rho}{Q}(0)=0$ and 0.01 the number of walls appears to be $\propto t^{-1}$, whereas for $\qsubrm{\rho}{Q}(0)=0.09$ and 0.25 it is clearly $\propto t^{-\gamma}$ where the scaling exponent $\gamma\ll1$. In fact for $\qsubrm{\rho}{Q}(0)=0.25$ the number of walls appears to freeze to a constant value at late times.

This is the main result of this paper. We find that domain walls formed in the presence of charge do not scale in the standard way and it appears that, if the initial charge density is sufficiently large, the configuration appears to freeze in. This is further illustrated by \fref{fig:freeze} which presents simulations with $\qsubrm{\rho}{Q}(0)=0.25$ and $\qsubrm{\rho}{Q}(0)=0.09$ beyond $t=\tau$ for $P=1024$, which was used in order for the simulations to be completed in an acceptable period of time. We see that, although the simulations contain a lower number of domains due to the smaller grid, there is very little evolution of the structures beyond $t=2\tau$.

In order to investigate the mechanism by which this non-standard scaling behaviour is achieved we have presented a number of different quantities for the simulation with $\qsubrm{\rho}{Q}(0)=0.09$ at $t=640$ in \fref{fig:quant}. We note that in each case we have tuned the colour scheme to highlight the important features of the plots. First in (a) we present the potential energy density, $\rho_{\rm PE}$. This is seen to be localized around regions corresponding to where $\phi$ goes from positive to negative confirming that domain walls have been formed. There is one very large loop which goes through a number of the periodic boundaries. Since it is possible to draw lines top-to-bottom and left-to-right which cross the boundaries but do not cross any of the walls, this loop does not wrap around the toroidal geometry imposed by the periodic boundaries. There are also a number of other smaller loops, most of which are clearly close to circular, and one which is in the middle of the plot which has a particularly non-regular shape.  

In (b), (c) and (d) we have presented $\Re{(\sigma)}$, $\qsubrm{\rho}{Q}$ and the modulus of the current $|{\bf J}|^2$. All three quantities are mainly localized around the wall structures seen in (a). We see that $\Re{(\sigma)}$ alternates from positive to negative as one traverses a particular loop, indicating that the $\sigma$-field is twisting along the wall; reminiscent of that seen in a kinky vorton. The fact that the charge and current are localized on the wall reinforces this conclusion.

We are, therefore, led to the conclusion that it is the charge and current on the walls which are counteracting the string tension in the same way as in kinky and cosmic vortons, and preventing the network from losing energy as fast as causality will allow as is necessary for the standard scaling regime. In fact, as we will discuss below, we believe that only two of the loops (the very large loop and the smaller unusually shaped loop in the middle of the plot) are close to stable, but the others are much longer-lived than they would be in the absence of charge.

In order to investigate whether the long-lived loops are indeed kinky vortons we have attempted to estimate the values of $N$, $Q$ and $R$ for some of the loops in the $\rho_Q(0)=0.09$ simulation. The loops we have selected are illustrated in \fref{fig:kinky}. They have been chosen since they are localized and present in the simulation when $t=640$ and $t=1280$. Two are close to being circular and the other is the unusually shaped loop already discussed above. The radius is estimated by calculating the area enclosed by the loop and equating this with $\pi R^2$; obviously this is only a crude approximation. Using the area is more accurate than the circumference since this is susceptible to errors induced by small-scale structure. The value of $Q$ is computed by including charge which is located within 10 grid squares from a point where a wall is identified. The value of $N$ is estimated by eye from \fref{fig:quant}(b), by counting the number of red to blue transitions. The values are presented in the table in \fref{fig:kinky}, along with the stable radius, $R_{\star}$, computed for the estimated values of $Q$ and $N$, using the formula given in ref.~\cite{BattyeSut2}.

We see that for loop 2 the values of $R$ and $R_{\star}$ are consistent with each other (within the somewhat rough calculation we have used for such a non-circular loop), and that the size of the loop reduces only a small amount with time. This indicates to us that this is close to a kinky vorton. For the other two loops (1 and 3) the estimated values of $R$ are somewhat larger than $R_{\star}$ and the values of $N$ have reduced quite a bit between the two timesteps. We conclude that the loops are not likely to stabilize, but that their collapse is being inhibited by the charge and current on the loops; however, one should note that the radii they are collapsing from is greater than the stable radii -- perhaps indicating that the loops could attain the stable radii, given sufficient simulation-time. We note that all the loops formed are in the electric regime, that is $Q \gg 4\pi N$.

It is difficult to reproduce a similar analysis for the case of $\rho_Q(0)=0.25$. Although there are actually many more loops, their density is much higher and they interact with each other, making it difficult to identify loops which are present at both timesteps. Moreover the equivalents of Figs.~\ref{fig:quant}(b),(c) and (d) are much more confusing and, in particular, it is difficult to estimate $N$ with any confidence for any loops. We have already pointed out that in this higher charge simulation the situation is much closer to that of dynamical frustration with the high density of walls leading to a quasi-stable network. However, true frustration will probably require the underlying field theoretic model to include junctions.

\section{Conclusions}

Our work has shown that the evolution of superconducting domain walls can lead to a network which does not necessarily scale in the standard way. In the case of a uniform background of charge as discussed in this paper, there appears to be a critical value of $\rho_Q(0)$ above which the scaling exponent $\gamma<1$ and possibly a higher charge where this becomes zero.

We have conclusively shown that charge and current are localized onto the loops and that they can be long-lived. Some appear to be close to stable and have the properties of kinky vortons. In the case of $P=4096$ and $\rho_Q(0)=0.09$ it was possible to synthesize just two loops which were close to stability and a host of others whose collapse was inhibited by the existence of current and charge. For lower values of $P$, for example 1024, there is only one large loop which is left at the end. We would presume that more stable loops could be found for larger values of $P$ and these small numbers are just a reflection of the low dynamic range.

Although kinky vorton solutions have been shown to exist and cosmic vortons have also been invesitgated in a narrow parameter range, their creation in a truely random scenario might have seemed a little implausible since they are dynamically, rather than topologically, stable. However, our conclusions appear to suggest that the formation of vortons as part of a more realistic scenario is at least possible. Of course, as we have seen most loops will be unstable and will ultimately collapse, but the cosmic vorton scenarios only require a very small number of loops to be absolutely stable. Unfortunately, to investigate the distribution of vorton sizes formed in a realistic scenario requires much greater dynamical range than that used in the current study.

Our simulations also appear to show possible direction for those interested in forming a frozen lattice from domain wall networks. For example, they form a caveat to the claims made in ref.~\cite{Avelino3} which state that dynamical frustration is impossible. Nonetheless, much work is necessary to turn this embryonic idea into a full theory of dark energy.
\section*{Acknowledgements}
We have benefited from code written by Chris Welshman which formed the basis of our visualization software. PMS acknowledges STFC for support under the rolling grant ST/G000433/1.

%\bibliographystyle{apsrev}
%\bibliography{kvform}

\end{document}